\documentclass[twocolumn]{aastex7}

\usepackage{amsmath}
\usepackage{hyperref}
\usepackage{multirow}

\usepackage{xspace}

\usepackage[range-phrase=-, range-units=single]{siunitx}

\hypersetup{pdfauthor={Aster Barry-Taylor}}
\begin{document}

\title{Measuring the Age and Initial Composition of Interstellar Objects}

 \author[0000-0002-0140-4475]{Aster G. Barry-Taylor}
 \altaffiliation{Fannie and John Hertz Foundation Fellow}
\affiliation{Dept. of Astronomy, University of Michigan, Ann Arbor, MI 48109, USA}
\email[show]{agtaylor@umich.edu}

\begin{abstract}

Measuring the age and composition of interstellar objects is necessary to understand their origins. However, existing methods for measuring the age are either imprecise or require an unusually bright target, while the composition is confounded by the effects of Galactic cosmic rays, which drive chemical reactions in the cometary matrix. As the Vera C. Rubin Observatory expands the census of interstellar objects, measuring these properties will be critical to understanding the overall population. In this work, I derive a new methodology for simultaneously estimating the age and initial composition of an interstellar object from time-series observations of the coma. Measurements of both the composition and its gradient versus depth into the nucleus are shown to fully constrain both properties in most circumstances. A simple model using only CO and CO$_2$ is discussed and used to demonstrate the properties of this inference process. When applied to 3I/ATLAS, this simple model indicates that this object is at least \qty{0.6}{Gyr} old and has an initial CO/CO$_2$ ratio of 1.5, supporting a formation outside the CO ice line. Further refinement of the chemical network and improved observational strategies will improve the precision and accuracy of these measurements, enabling a deeper understanding of the interstellar object population and extrasolar planet formation. 

\end{abstract}


\section{Introduction} \label{sec:intro}

Although planet formation occurs throughout the Galaxy, the distance sensitivity of planet detection methods means that our probes of this process are limited to the Solar neighborhood. Since interstellar objects are products of the planet formation process and are well mixed throughout the Galaxy, they uniquely probe planet formation across Galactic space and history. While only three interstellar objects --- 1I/`Oumuamua \citep{MPEC2017, Williams2017}, 2I/Borisov \citep{Borisov2019}, and 3I/ATLAS \citep{Denneau2025, Seligman2025} --- have been discovered to date, the Vera C. Rubin Observatory's Legacy Survey of Space and Time is expected to find \numrange{10}{100} more interstellar objects during its \qty{10}{yr} mission \citep{Hoover2022, Marceta2023, Dorsey2025}. This sample will be a remarkable opportunity to study planet formation over broad scales.

Two particular properties of interstellar objects --- their compositions and their ages --- provide this insight. The composition of these objects traces the chemistry of extrasolar environments, while the age traces the Galactic history of planet formation. The age distribution of interstellar objects can be used to infer their formation rate. The ratio between the interstellar object formation rate and the Galactic star formation rate is then the number of interstellar objects produced per star throughout Galactic history \citep{Taylor2025}. This production efficiency is set by the formation of both planetesimals and widely separated giant planets, which create interstellar objects through migration (see, e.g., \citealt{Laughlin2017}). Although these effects are degenerate, this connection means that the interstellar object age distribution is a unique probe of exoplanet formation and dynamics. 

However, these properties are difficult to measure. Although the coma composition of some interstellar objects can be measured directly, this composition need not reflect an object's formation environment. Interstellar objects, by definition, have spent a significant amount of time in the interstellar medium outside of the bubble carved by any stellar wind, where they are subject to the effects of Galactic cosmic rays (GCRs). Both theoretical models \citep{Gronoff2020, Maggiolo2020, Bergner2023, Maggiolo2026} and laboratory experiments (e.g., \citealt{Johnson1997, Hudson1999, Dartois2015}) show that GCRs modify the surface abundance of these objects from their primordial value, which is reflected in their coma compositions. Changes in the coma composition of interstellar objects \citep{Bannister2020, Aravind2021, SalazarManzano2025} and solar system comets \citep{AHearn1995, Pinto2022} during perihelion passage provide strong evidence of this effect. As a result of this processing, the coma compositions of interstellar objects do not directly reveal their initial composition. While post-perihelion compositions are often assumed to be primordial, this is not necessarily the case (see, e.g., \citealt{Maggiolo2026, Frincke2026}).

The age of interstellar objects is also difficult to infer. The inherently chaotic behavior of their Galactic trajectories means that it is impossible to trace an interstellar object back to its primordial system \citep{Hallatt2020}. As a result, the (unknown) parent star's age cannot be used as a proxy for the age of an interstellar object. While methods exist for measuring interstellar object ages, they are limited. For example, since stars (and interstellar objects) are born in the Galactic plane and then dynamically heated, the velocity of an interstellar object is a proxy for its age. While measurements of the ages of interstellar objects using their kinematics can provide useful information, the kinematic age posterior is broad (with \qty{1}{\sigma} bounds of a few \unit{Gyr}; \citealt{Hopkins2025, Taylor2025}) and therefore difficult to use to probe planet formation rates. For 3I/ATLAS, the $^{12}$C/$^{13}$C carbon isotope ratio in combination with Galactic chemical evolution constrains the age more precisely than the kinematic method \citep{Cordiner2026}. However, measuring the isotope ratio requires a relatively bright interstellar object with an active coma, which is not guaranteed. 

Current age measurement techniques are therefore complementary. While kinematic ages can be calculated for every interstellar object, these ages are imprecise. On the other hand, isotope ratios provide high-precision age measurements, but are limited to a small subset of large and bright interstellar objects. It also remains difficult to measure these objects' initial compositions. In this work, I introduce a methodology to simultaneously measure these properties. This technique leverages the fact that the amount of GCR processing decreases with depth, meaning that the composition and composition gradient (usually) uniquely specify the initial composition and age. This method generalizes previous work that used the onset of molecular emission to estimate the age using GCR processing \citep{Rawat2026a}. Although this method would benefit from theoretical work to improve precision, it can be applied to any interstellar object with a detectable coma, falling into a regime between the other two measurement techniques.  

This paper is structured as follows: in Sec.~\ref{sec:CRproc} I introduce the model in general terms and discuss its limitations. To build intuition, Sec.~\ref{sec:simpnet} then demonstrates the use of this method for a simple two-molecule chemical network. This network is then applied to 3I/ATLAS in Sec.~\ref{sec:application}. Finally, this paper concludes in Sec.~\ref{sec:conc}. 

\section{Processing by Cosmic Rays}\label{sec:CRproc}

This section studies the effects of cosmic rays on the composition of a cometary body. In the general case, consider a spherical comet with radius $R$ composed of $N$ molecular species. 
These species are fully arbitrary and can be any molecule (such as H$_2$O or CO) or structural state (amorphous versus crystalline ice). The composition of the comet at a depth $z$ from the surface is then characterized by an $N$-dimensional vector $\vec{X}(z)$, defined such that the sum of all components of $\vec{X}$ is the number density of the body. 

\subsection{General Solution}

The comet is assumed to have a homogeneous composition before ejection, which is then modified by the GCRs. Assuming that diffusion through the cometary matrix is small, the cosmic rays drive a local chemical network of the $N$ molecular species.
The behavior of this chemical network can be described by a rate matrix $\mathbf{A}$ and a vector $\vec{B}$, so that 
\begin{equation}\label{eq:basicODE}
    \frac{\partial}{\partial t}\vec{X}=\mathbf{A}\Phi\vec{X}-\text{diag}(\vec{B})\Phi\vec{X}+\mathcal{O}(\vec{X}\otimes\vec{X})+...
\end{equation}
In this equation, $\Phi$ is the effective cosmic ray flux and $\otimes$ is the tensor product. The matrix $\mathbf{A}$ captures the conservative conversion of one molecule into another, while $\vec{B}$ represents loss from the system. While Eq. \eqref{eq:basicODE} is fully generic, this work ignores the higher-order terms, which account for chemical reactions that require more than one reactant, for example, 
\begin{equation*}
\text{CO}+\text{O}\rightarrow \text{CO}_2\,.
\end{equation*}
Such a reaction has a rate that depends on both the abundance of CO and O, and hence cannot be represented by a linear term. In general, I assume that these reactions are fast or limited by a single reactant, so that they can be well approximated by the linear coefficient $\mathbf{A}$. 

Ignoring the $\vec{X}\otimes\vec{X}$ terms, Eq. \eqref{eq:basicODE} can be solved for $\vec{X}(t)$ given an initial condition of $\vec{X}(0)=\vec{X}_0$. Let the $N\times N$ matrix $\mathbf{R}=\mathbf{A}-\text{diag}(\vec{B})$, which is assumed to be fixed in time.
The general solution to Eq.~\eqref{eq:basicODE} then takes the form 
\begin{equation}\label{eq:gensol}
    \vec{X}(t)=\exp[\mathbf{R}\Phi t]\vec{X}(0)\,.
\end{equation}
If $\mathbf{R}$ is diagonalizable with eigenvectors $\vec{v}_i$ and eigenvalues $\lambda_i$, then this equation reduces to
\begin{equation}\label{eq:diagX}
    \vec{X}(t)=\sum_{i=1}^Nc_i\exp(\lambda_i t)\vec{v}_i\,.
\end{equation}
The constants $c_i$ are defined to be the components of the initial vector $\vec{X}(0)$ in the eigenbasis of $\mathbf{R}$. 

Since $\mathbf{A}$ represents only conservative conversions, its diagonal entries satisfy $A_{ii}=-\sum_{j\neq i}A_{ji}$, that is, the diagonal entries in the matrix are equal to the negative sum of their corresponding off-diagonal column. As a result, the row vector $\vec{1}^\top$ vector is a left-eigenvector with an eigenvalue of zero. The matrix $\mathbf{A}$ is therefore singular and has a right-eigenvector with zero eigenvalue. This eigenvector is a steady state of the chemical network. 

Assume that the network represented by $\mathbf{A}$ cannot be decomposed into simpler, independent networks. By the Perron-Frobenius theorem, there is only one steady state, and the remaining eigenvalues must have a real part $\Re[\lambda_i]<0$ (see Appendix~\ref{app:negeigproof} for a proof of this claim). While the imaginary component can be nonzero (causing rotation of the state), the real part being negative means that the magnitude of all other states will decay. In this situation, the composition exponentially decays to the steady state (perhaps with some rotational component), with the decay rate of each eigenvector given by the real component of the eigenvalue. 

In physical terms, the (irreducible) conservative network is required to converge to some steady state. If there is more than one steady state $\vec{v}_{0,k}$, then the chemical network can be broken down into two independent components, each of which will decay to their steady states in turn. If a nonzero $\vec{B}$ is included, then unless the steady state(s) have no loss terms, the total abundance drops to zero over time. In most realistic circumstances, the eigenvalues of $\text{diag}(\vec{B})$ are much smaller than the eigenvalues of $\mathbf{A}$, so that the network relatively quickly decays to a steady state that then slowly decays over time (e.g., \citealt{Pilling2022}).  

In addition to being a function of time, the composition is also a function of depth into the comet. In particular, the principal driver of these network reactions is the cosmic ray flux. After all, these reactions are driven by the GCR flux $\Phi(z)$, which is a function of depth. For the sake of definiteness, I will assume that the cosmic ray flux at a given depth $z$ is independent of time (see, e.g., \citealt{Smith2019}).

\subsection{Depth Dependence}

The depth dependence of $\Phi$ has important implications. Measurements of the outgassing comae of interstellar objects in the solar system commonly reveal the composition (i.e., the values of $\vec{X}$) at some depth. At the same time, the observed time evolution of the composition can reveal the gradient of the composition with depth into the interstellar object. 
When the surface of the interstellar object is processed by sublimation, the escaping material samples progressively deeper layers. Observations of the composition during this process can measure the composition rate of change $\partial\vec{X}/\partial t_{\rm obs}$. Given a model for the sampling depth over time $\partial z/\partial t_{\rm obs}$ (either due to thermal processing or lost mass, see \citealt{Yaginuma2026, Frincke2026}, and Sec.~\ref{sec:application}), this changing composition can be directly related to the compositional gradient via $\partial\vec{X}/\partial z=(\partial\vec{X}/\partial t)(\partial t/\partial z)$.
Observations can thus provide $\partial\vec{X}/\partial t$ and $\vec{X}$, while $\partial \Phi/\partial z$ and $\mathbf{R}$ are theoretically known. The combination of these parameters completely specifies the age and initial composition of the object.

To demonstrate this, first consider the depth derivative of the composition vector $\vec{X}$. Differentiating Eq.~\eqref{eq:gensol} and assuming that $\mathbf{R}$ and $\vec{X}_0$ are independent of depth shows that
\begin{equation}\label{eq:Xgrad}
    \frac{\partial}{\partial z}\vec{X}=\frac{\partial\Phi}{\partial z}\mathbf{R}t\exp[\mathbf{R}\Phi t]\vec{X}_0=\frac{\partial\Phi}{\partial z}\mathbf{R}t\vec{X}\,.
\end{equation}
With all components known, this equation can be solved to find the age $t$. The initial composition $\vec{X}_0$ is then easily found with
\begin{equation}
    \vec{X}_0=\exp[-\mathbf{R}\Phi t]\vec{X}\,.
\end{equation}
Measurements of the composition and the composition gradient thus provide the age and the initial composition. 

Note that this setup relies on knowing $\Phi(z)$, and therefore requires good knowledge of the depth into the comet that is sourcing the emission. While this ``sampling depth'' can be independently calculated from a time series of dust and gas mass loss rates, its value is highly uncertain and can drive significant uncertainty in the derived parameter values. 

Note that if $\partial\vec{X}/\partial z$ is not parallel to $\mathbf{R}\vec{X}$, then there is no value of $t$ that satisfies the equation. The severity of this issue depends on the magnitude of the discrepancy. While misalignment is expected due to both observational and model uncertainties, significant misalignment would indicate that the model used is seriously flawed. 
There can also be serious issues if $\partial\vec{X}/\partial z$ is zero or close to it. In this case, all that can be learned about the age and the initial condition is that $\vec{X}$ is close to the null space of $\mathbf{R}$, or that $t$ is sufficiently large as to overwhelm the non-null-space component. Effectively, if the composition is flat with depth, then the composition must either have started near the steady state or be sufficiently old that the steady state has been reached. 


\subsection{Additional Parameter Constraints}

For a network with $N$ chemical species, there are $N+1$ unknown parameters ($\vec{X}_0$ and $t$).\footnote{Note that there may be somewhat exotic molecules in the network that have safely assumed initial values. If there are $M$ such molecules, then there are only $N-M+1$ true unknowns.} Assume that the mass loss rates of $J\leq N$ species are observed, along with the changing mass loss rate over time (hence, through some modeling, the composition gradient with depth). These observations provide $2J$ observables for the model. The remaining parameters are assumed to be known, through direct measurement (e.g., the radius $R$), standard values (the density $\rho$), or rough approximation (the Bond albedo $A$). It is especially critical that the chemical network parameters $\mathbf{A}$ and $\vec{B}$ are at least approximately known, since there are $N^2$ parameters in this network in total. 

In this circumstance, constraining the initial composition and age requires that $J\geq (N+1)/2$, so that for large $N$ at least half of the species must be observed. The addition of the composition gradient is a major improvement, since without these measurements the constraint becomes $J\geq N+1$. Since $J\leq N$, this problem is never fully constrained without the composition gradient. If more species than necessary are observed, then this problem can be overconstrained. Although overconstraining presents some difficulty, it also presents an opportunity. A complete model of this process requires a number of variables --- the age of the interstellar object $t$, its initial radius $R_0$, density $\rho$, heat capacity $C_p$, thermal conductivity $\kappa$, Bond albedo $A$, emissivity $\epsilon$, the initial composition $\vec{X}_0$, the chemical network parameters $\mathbf{A}$ and $\vec{B}$, etc. An overconstrained problem allows for some of these parameters to freely vary and be constrained as well. In general, if half these parameters can be known or measured, then the rest can be uniquely determined.

For $J=N$, every additional species in the network provides a constraint on another free parameter. Given that there are only approximately seven additional parameters that may be freed in this simple model (e.g., $\rho$, $C_p$, $\kappa$, $\epsilon$, $A$, $R_0$, and the $e$-folding depth of the cosmic ray flux $H$), an eight-species network with complete observational coverage fully specifies the system. For 3I/ATLAS, nine species have been observed over an extended arc (see \citealt{Frincke2026}), demonstrating that such observations are feasible, if difficult. 


\section{Simple Example Network}\label{sec:simpnet}

While chemical networks from cosmic ray processes have not reached the maturity necessary to deploy this technique to its full potential, I will demonstrate the applicability of this methodology with a simple two-molecule network --- CO and CO$_2$, which are the dominant targets in previous laboratory studies (e.g., \citealt{Hudson1995, Moore1998, Pilling2010}) of the effects of ion irradiation on astrophysical ices. 

Before further discussion, some comments must be made about the parameters used. These reactions are each characterized by a single number that describes the reaction rate. In this work, this number is not the geometric cross-section $\sigma$ (in units of \unit{cm^2}), which represents the interaction probability for a single incoming particle. The issue here is the energy dependence of the cross-section, which can be dramatic. As the GCRs travel through the interstellar ice, their energy spectrum and effective cross-section change. Fortunately, chemical reaction rates are relatively insensitive to the energy or species of the impinging particle. Instead, these reactions are driven almost entirely by the amount of energy deposited into the relevant ice. Instead of tracking the cross-section $\sigma$ and the particle flux, I therefore track the reaction rate $k$ (in units of \unit{cm^3/eV}) and the energy dissipation rate $\Phi$ (in \unit{eV/cm^3/s}). The form of the solutions discussed in Sec.~\ref{sec:CRproc} is unchanged, but the network parameterizations require some additional work. 


\citet{Pilling2010} studied the effects of irradiation by \qty{52}{MeV} Ni ions onto ice of various mixtures of H$_2$O and CO$_2$. This work found that (within an order of magnitude) the total destruction rate of CO is $k_1\simeq\qty{e-23}{cm^3/eV}$ and the total destruction rate of CO$_2$ is $k_2\simeq\qty{e-24}{cm^3/eV}$. These values are found by fitting the CO/CO$_2$ ratio versus fluence data (Fig. 7 in \citealt{Pilling2010}), using the stopping rates reported in the text. Comparing the reported production cross-section of CO to the destruction cross-section of CO$_2$ reveals that almost all of the destroyed CO$_2$ is transformed into CO. While the production of CO$_2$ is not studied separately, inspection of the results of \citet{Hudson1999} shows that most of the CO is in turn transformed into CO$_2$ and confirms this value of the rate coefficient, although there is $\mathcal{O}(1)$ disagreement between these studies. 
This disagreement is likely the result of measurement imprecision, in particular the \qty{100}{\percent} error in the sample thicknesses reported by \citet{Hudson1999}.

\citet{Pilling2010} also reports that the CO/CO$_2$ ratio approaches a steady state of approximately \num{0.1} before the abundance of both molecules begins to decay. This ratio is independent of the composition of the ice, and the rate coefficients change by factors of only a few across compositional ratios. Similar experiments (e.g., \citealt{Moore2000, Strazzulla2005, Pilling2022}) also show that the rate coefficient $k$ in this form is roughly constant across ice compositions. 

For the sake of simplicity, in this model the loss-rate term $\vec{B}$ is set to zero. While the laboratory experiments show decay, the ice samples are very thin ($\lesssim\qty{1}{\micro\meter}$) and these decay rates are mostly due to sputtering. The decay is likely to be much less significant in the case of an interstellar comet, where there is more material impeding molecular escape. The rate matrix for the network is then taken to be
\begin{equation}
    \mathbf{R}=\begin{bmatrix} 
                    \num{-e-23} & \num{e-24}\\
                    \num{e-23} & \num{-e-24}
                    \end{bmatrix}\unit{cm^3/eV}\,.
\end{equation} 
Since this rate matrix is diagonalizable, Eq.~\eqref{eq:diagX} is used. For this rate matrix, the steady-state composition ratio is $\beta\equiv n_{\rm CO}/n_{\rm CO_2}=0.1$, or $\vec{v}_1=[0.1, 1]$. The other eigenvalue is $-k\equiv-k_1-k_2=\qty{-1.1e-23}{cm^3/eV}$, with an eigenvector of $\vec{v}_2=[-1, 1]$. Given these values, the evolution of $X_{\rm CO}$ and $X_{\rm CO_2}$ is simple --- for initial values of $X_{\rm CO,0}$ and $X_{\rm CO_2,0}$ and a total CO and CO$_2$ density $X_t$, the time-dependent solutions take the form
\begin{subequations}
\begin{align}
\begin{split}\label{eq:XCO}
     X_{\rm CO}(z,t)=&X_{\rm CO,0}\exp[-k\,\Phi(z)t] \\
     &+ \frac{k_2}{k}X_t\left(1-\exp[-k\,\Phi(z)t]\right)\,,\\
\end{split}\\
\begin{split}\label{eq:XCO2}
    X_{\rm CO_2}(z,t)=&X_{\rm CO_2,0}\exp[-k\,\Phi(z)t] \\
    &+ \frac{k_1}{k}X_t\left(1-\exp[-k\,\Phi(z)t]\right)\,.\\
\end{split}
\end{align}    
\end{subequations}
That is, the initial abundance decays exponentially and transforms into the steady-state abundance with a decay rate of $k\Phi(z)$. It is worth noting that since $X_t$ is conserved in this case, the ratio between the CO and CO$_2$ abundances depends only on the initial abundance ratio, the steady-state abundance ratio, and $k\Phi(z)t$. For simplicity and to cancel the uncertainty regarding the abundance with respect to water ice, I here study the CO/CO$_2$ ratio (called $\beta$) rather than the abundances alone. While some information is lost in this simplification, the resulting mathematics is easier to understand. This ratio also corrects for variations in the total activity during perihelion passage. The time- and depth-dependent ratio $\beta(z,t)$ is then given by 
\begin{equation}\label{eq:betaval}
    \beta(z,t)=\frac{\beta_{\rm eq}+\gamma \exp[-k\,\Phi(z)t]}{1-\gamma \exp[-k\,\Phi(z)t]}\,,
\end{equation}
where $\beta_{\rm eq}=k_2/k_1$ is the steady-state ratio. The decaying component $\gamma$ is given by 
\begin{equation}
    \gamma=\frac{\beta_0-\beta_{\rm eq}}{1+\beta_0}\,,
\end{equation}
where $\beta_0$ is the initial value of $\beta$. Replacing $\gamma$ in Eq.~\eqref{eq:betaval} and expanding confirms that this form is equal to the ratio of Eqs.~\eqref{eq:XCO} and \eqref{eq:XCO2}.

For now, the energy dissipated by cosmic rays $\Phi(z)$ is assumed to take the form of an exponential with a normalization $\Phi_0$ and an $e$-folding depth of $H$, so that
\begin{equation}
    \Phi(z)=\Phi_0\exp(-z/H)\,.
\end{equation}
For cosmic rays, the $e$-folding depth is approximately $H\simeq\qty{4}{m}$ for depths of $z\lesssim\qty{35}{m}$ into the comet's surface (see \citealt{Gronoff2020, Maggiolo2026}). Beyond this depth, the $e$-folding depth increases as the principal cosmic ray source becomes secondary particles. The normalization of the energy deposition rate is taken to be $\Phi_0=\qty{1.5e7}{eV\,cm^{-3}\,s^{-1}}$ \citep{Gronoff2020}. In Sec.~\ref{sec:application}, this depth dependence is interpolated from simulated results, so this functional form is merely illustrative. 

\begin{figure}
    \centering
    \includegraphics{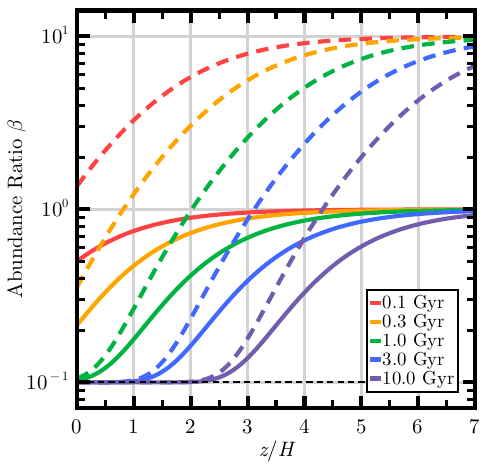}
    \caption{\textbf{Evolution of CO/CO$_2$ Ratio.} The CO/CO$_2$ ratio $\beta$ for different initial compositions and ages versus depth. The solid lines have $\beta_0=1$ and the dashed lines $\beta_0=10$. The colors indicate different ages. The black dashed line shows the equilibrium value of $\beta_{\rm eq}=0.1$. }
    \label{fig:ratevo}
\end{figure}

Note that the total density of CO and CO$_2$ is constant, so that only the ratio varies as a function of time. Fig.~\ref{fig:ratevo} shows the CO/CO$_2$ ratio as a function of time and depth into the comet for initial ratios of $\beta_0=1$ and $\beta_0=10$. This figure shows that the abundance ratio and gradient evolve over time, converging to the steady-state value of $\beta=0.1$. For a comet with $t\geq\qty{1}{Gyr}$ and these initial conditions, the surface abundance and slope are essentially uninformative and flat. In this instance, measurements of the composition and slope only inform about the value of the steady state and place a degenerate joint constraint on $\beta_0$ and $t$ . 

\begin{figure}
    \centering
    \includegraphics{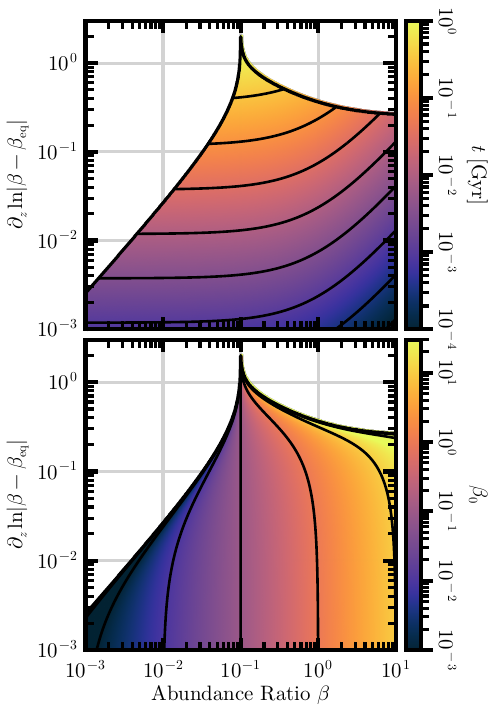}
    \caption{\textbf{Age and Initial Abundance vs. Measured Parameters.} This figure shows the age $t$ (top panel) and initial abundance $\beta_0$ (bottom panel) that would be derived from a measurement of a certain abundance ratio $\beta$ and gradient $\partial/\partial z (\ln|\beta-\beta_{\rm eq}|)$. The white regions are combinations of parameters that are not possible within this model (i.e., the initial abundance is negative). }
    \label{fig:ratiocalc}
\end{figure}

In most circumstances, the ratio and ratio gradient at a given depth uniquely provide the age and initial ratio of an object. Fig. \ref{fig:ratiocalc} shows the age $t$ and initial CO/CO$_2$ ratio $\beta_0$ implied by a given measured surface CO/CO$_2$ ratio and ratio gradient. In order to meaningfully visually sample the parameter space, the gradient shown is not $\partial\beta/\partial z$ directly but $\partial/\partial z (\ln|\beta-\beta_{\rm eq}|)$, where $\beta_{\rm eq}$ is the equilibrium CO/CO$_2$ abundance ratio. This value corresponds directly to $\partial\beta/\partial z$. 

Fig. \ref{fig:ratiocalc} has several implications. First, some combinations of $\beta$ and $\partial\beta/\partial z$ are forbidden. Since the gradient and ratio are tied together, these regions can only be achieved if $\beta_0<0$, which is physically impossible. Measurements in such a forbidden region would therefore indicate that the model used is overly simplistic. 

Second, at $\beta$ values close to $\beta_{\rm eq}$, the problem becomes degenerate. This is the cause of the spike at $\beta=0.1$. Note that this pole is a valid solution for all initial values $\beta_0$, although it converges at a range of ages. Therefore, if the spatial gradient of the ratio is small, then the measurement only puts a joint constraint on $\beta_0$ and $t$, rather than individual constraints. A measurement of a vanishing gradient would also demonstrate that the measured CO/CO$_2$ ratio in the interstellar object is the equilibrium value, which would be useful for constraining the effects of cosmic ray processing.

Finally, it is important to note that this figure reports only the surface abundance of CO/CO$_2$. In a real comet, erosion has removed some of the surface, and so the measured values sample some depth into the surface. Knowing this depth is critical for calibrating the derived results. However, a similar figure can be constructed for any known depth into the object and provides useful intuition. In addition, the CO and CO$_2$ outgassing can originate from different depths in the nucleus, which can further complicate the ratio comparison.

\begin{figure}
    \centering
    \includegraphics{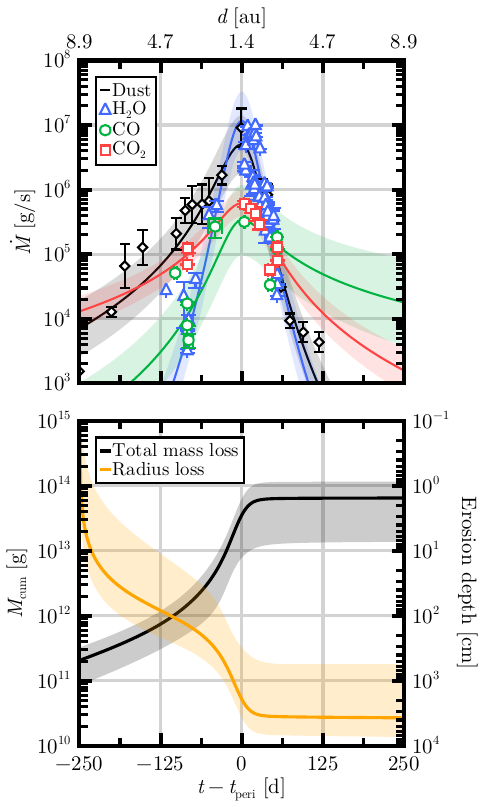}
    \caption{\textbf{Mass and Radius Loss of 3I/ATLAS.} Observational data of the mass loss rates of dust, H$_2$O, CO, and CO$_2$ (top panel) of 3I/ATLAS. Power-law curves are also fit to these observations (lines). The cumulative mass loss and the change in radius due to this mass loss (assuming a \qty{3}{\percent} active fraction and a \qty{1.3}{\kilo\meter} nucleus) are shown in the bottom panel. }
    \label{fig:massloss}
\end{figure}

\section{Application to 3I/ATLAS}\label{sec:application}

As a proof of concept, in this Section I apply this methodology to the third interstellar object 3I/ATLAS. The CO/CO$_2$ ratio of this object has been measured several times during its passage through the solar system, both pre- and post-perihelion. These measurements allow for an approximate calculation of the age and initial composition of this object using this technique. I assume that the evolution of the CO/CO$_2$ ratio is described by the simple two-molecule network described in Sec.~\ref{sec:simpnet}. However, it is difficult to accurately model the activity of CO and CO$_2$ separately without a full thermal model (left for later work). Instead, I study the CO/CO$_2$ ratio, which cancels out many of the complicated factors.

The sampling depth --- the depth of the source of the outgassing material relative to the primordial surface --- is needed to make use of the CO/CO$_2$ ratio to derive the age of 3I/ATLAS. This depth is derived from the cumulative mass loss of the object, since a given density $\rho$ and total mass loss $M$ correspond to a volume, which then sets the surface erosion. The geometry of the mass loss, which converts from the volume to the change in radius, is somewhat complicated. Previous work on this topic \citep{Frincke2026} has assumed that the lost mass is removed from a uniform layer of the surface, although anisotropic mass loss is common for comets in general (e.g., \citealt{AHearn1995}) and 3I/ATLAS in particular \citep{Serra-Ricart2025, Jewitt2025b}. 

I instead assume that a fraction $f_A$ of the object's surface is active and that all activity originates from this region. Under this model, the sublimating region is at the bottom of one or more pits, which are carved out by the material loss. Given a mass loss rate $\dot{M}$, a spherical nucleus with a radius $R_N$, and an active fraction $f_A$, the time derivative of the depth of the sublimation region $z_s$ is 
\begin{equation}
    \dot{z}_s=\frac{\dot{M}}{4\pi R_N^2\rho f_A}\,.
\end{equation}

To find the total mass loss, power laws are fit to the observed outgassing rates of CO, CO$_2$, H$_2$O, and dust and the fits are integrated in time. The molecular outgassing rates are collated in \citet{Frincke2026}, while the dust data are taken from \citet{Moreno2026}. While two separate power laws are used for pre- and post-perihelion passage, these fits are forced to have the same value at perihelion itself, which ensures that the mass loss is continuous. The optimal parameters are found using the weighted linear least-squares algorithm in log space. The measurement uncertainties are extrapolated into parameter uncertainties, with an irreducible minimum uncertainty from the measurement scatter.

The mass loss rates, best-fit curves, and confidence regions are shown in the top panel of Fig.~\ref{fig:massloss}. This figure shows that CO$_2$ dominated the pre-perihelion outgassing, while H$_2$O dominated post-perihelion. The total, cumulative mass loss (found by integrating the individual loss rates) is also shown in the bottom panel of this figure. 

For 3I/ATLAS, the extreme hyperactivity of H$_2$O complicates finding the active fraction. This hyperactivity is the result of H$_2$O sublimation from icy coma grains, as confirmed by the dependence of H$_2$O production on telescope aperture \citep{Cordiner2025, Bockelee-Morvan2026}. On the other hand, the CO and CO$_2$ emission appears to be localized to the nucleus, indicating primary emission from an active nucleus \citep{Cordiner2025, Cordiner2026}. Solving the energy balance for the observed $Q$ values using the physical constants for CO and CO$_2$ \citep{Brown1980} and hemispheric radiation yields a CO$_2$ active fraction of approximately \qty{3}{\percent} (assuming a radius of $R_N\simeq\qty{1.3}{km}$; \citealt{Hui2026}). The CO active fraction is smaller until perihelion, at which point the CO$_2$ active fraction drops and the CO active fraction rises to \qty{2}{\percent}. Based on these results, a constant active fraction of $f_A\simeq\num{0.03}$ is used here. 

The depth of the active region using this value of $f_A$ is also shown in the bottom panel of Fig.~\ref{fig:massloss}. Long after perihelion, the total radius loss ranges from \qtyrange{6}{70}{m}, with significant uncertainty from the CO mass loss fit. The differences between this range and the \qtyrange{1.7}{5.2}{m} from \citet{Frincke2026} are ascribed to (i) the inclusion of dust data, (ii) the enforcement of continuity at perihelion, and (iii) the non-isotropic active fraction. Note that this radius loss may be larger than the \qty{35}{m} over which an exponential decay is appropriate. The energy dose versus depth from cosmic rays for 3I/ATLAS is therefore calculated by interpolating the dose profile of \citet{Maggiolo2026} rather than using a simple exponential decay. 

\begin{table*}
\centering
\caption{\textbf{CO/CO$_2$ Ratio Measurements.} Directly measured observations of the 3I/ATLAS CO/CO$_2$ production-rate ratio.}\label{tab:coco2rats}
\begin{tabular}{lccllccl}
Epoch & $t - t_{\rm peri}$ & $r_h$ & Facility & $Q_{\rm CO}$ & $Q_{\rm CO_2}$ & $Q_{\rm CO}/Q_{\rm CO_2}$ & Reference \\
(2025) & [d] & [au] & & [\qty{e26}{s^{-1}}] & [\qty{e26}{s^{-1}}] & & \\
\hline
Aug 6 & $-84$ & 3.32 & JWST/NIRSpec & $3.7 \pm 0.2$ & $17.0 \pm 0.1$ & $0.218 \pm 0.012$ & \citet{Cordiner2025} \\
Aug 1--15 & $-82$ & 3.5--3.1 & SPHEREx & $1.0 \pm 0.25$\textsuperscript{a} & $16 \pm 1.6$ & $0.063 \pm 0.017$ & \citet{Lisse2026} \\
\multirow{2}{*}{Nov 1--3} & \multirow{2}{*}{$+4$} & \multirow{2}{*}{1.36} & IRAM 30-m & $68.0 \pm 1.1$ & \nodata & \multirow{2}{*}{$0.83 \pm 0.08$\textsuperscript{b}} & \citet{Biver2026} \\
 &  &  & JUICE/MAJIS & \nodata & $82 \pm 8$ &  & \citet{Bockelee-Morvan2026} \\Dec 8--15 & $+43$ & 2.0--2.2 & SPHEREx & \nodata & \nodata & $0.92$\textsuperscript{c} & \citet{Lisse2026a} \\
Dec 22--23 & $+54$ & 2.37--2.42 & JWST/NIRSpec & $39.4 \pm 0.1$ & $17.6 \pm 0.1$ & $2.24 \pm 0.01$ & \citet{Cordiner2026} \\
Dec 22--23 & $+54$ & 2.37--2.42 & JWST/NIRSpec & $25.5\pm0.2$ & $10.8\pm0.2$ & $2.36 \pm 0.05$ & \citet{Roth2026a} \\

\end{tabular}
\begin{flushleft}
\footnotesize
Note: Epochs spanning a range list the midpoint for $t - t_{\rm peri}$. 
\textsuperscript{a}Tentative CO detection. 
\textsuperscript{b}Cross-instrument pairing at a shared epoch. 
\textsuperscript{c}No uncertainties or direct $Q$ values are published.
\end{flushleft}
\end{table*}

\begin{figure}
    \centering
    \includegraphics{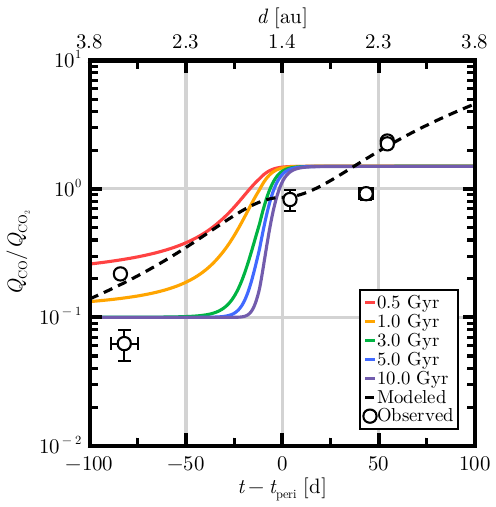}
    \caption{\textbf{CO/CO$_2$ Ratio of 3I/ATLAS.} Observational measurements of the CO/CO$_2$ ratio of 3I/ATLAS, along with the CO/CO$_2$ ratio derived by extrapolating the outgassing rate models shown in Fig.~\ref{fig:massloss} (black). The expected CO/CO$_2$ ratio profiles for a 3I/ATLAS model with $\beta_0=1.5$ and a range of ages are also shown.}
    \label{fig:coco2ratio}
\end{figure}

Using this active region depth over time and the CO/CO$_2$ ratio versus depth calculated from Eq.~\eqref{eq:betaval}, the CO/CO$_2$ ratio over time can be predicted from an age and an initial CO/CO$_2$ ratio. Some model curves for 3I/ATLAS using $\beta_0=2$ and a range of ages are shown in Fig.~\ref{fig:coco2ratio}. In addition to the extrapolated CO/CO$_2$ ratio, there are also six direct measurements of the CO/CO$_2$ ratio of this object. These points are shown in Fig.~\ref{fig:coco2ratio} and are collected in Table~\ref{tab:coco2rats}. While there is significant scatter in the observed data --- likely due to different measurement techniques, instruments, and reduction methods --- the broad shape of the theoretical curve is a reasonable match for these observations.

\begin{figure}
    \centering
    \includegraphics{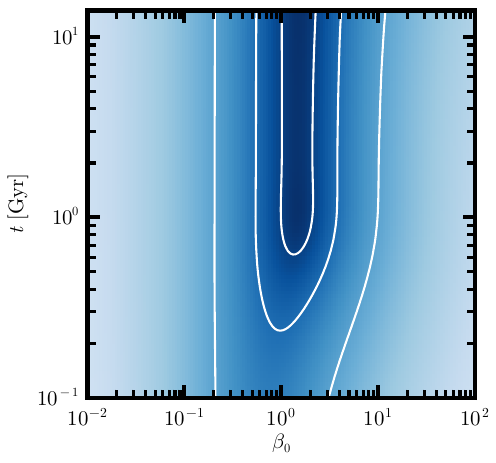}
    \caption{\textbf{Model Likelihood Values.} The log-likelihood values for a CO/CO$_2$ model fit to the observational data points versus initial compositions $\beta_0$ and age $t$. The 1, 2, and \qty{3}{\sigma} contours are shown in white. The optimal values are $\beta_0\simeq1.5$ and $t\gtrsim\qty{0.6}{Gyr}$.}
    \label{fig:modelchi2}
\end{figure}

Due to the speed of this model, it is relatively trivial to calculate the CO/CO$_2$ ratio over time for a given initial ratio and age while keeping the behavior of the chemical network and the sampling depth versus time fixed. 
I calculate the relative likelihood for the model given the data over a grid of $\beta_0$ and $t$ values. This calculation also includes a dummy model error parameter $\sigma_{\rm int}$. The resulting likelihoods (marginalized over $\sigma_{\rm int}$ are shown in Fig.~\ref{fig:modelchi2}. 
The priors for $\beta_0$, $t$, and $\sigma_{\rm int}$ are log-uniform over $\beta_0\in[\num{e-2}, \num{e2}]$, $t\in[\num{e-2}, \num{14}]\,\unit{Gyr}$, and $\sigma_{\rm int}\in[\num{e-2},\num{e1}]$. 
This model error parameter is constrained to $\ln\sigma_{\rm int}\simeq\num{0.62(0.31:0.17)}$. The likelihood peaks at $\beta_0=\num{1.50(0.55:0.41)}$ and is flat for $t\gtrsim\qty{0.6}{Gyr}$, where the quoted uncertainties are the \qty{1}{\sigma} confidence intervals.

This limit is consistent with the ages calculated via kinematics \citep{Taylor2025, Hopkins2025} and isotopic ratios \citep{Cordiner2026}, while this method simultaneously estimates the initial CO/CO$_2$ ratio of 3I/ATLAS. This ratio is within the typical range for comets \citep{AHearn2012, Pinto2022}, but suggests that 3I/ATLAS formed in the outer edges of its natal disk. Although these estimates have large error bars, these errors can be reduced through careful theoretical work and observation campaign designs for future interstellar objects. I discuss these possibilities in detail in Sec.~\ref{sec:conc}.

\section{Conclusion}\label{sec:conc}

\subsection{Summary}

In this work, I have presented a novel methodology for estimating the age and initial compositions of interstellar objects using GCR processing of the objects' surface. This method relies on measuring both the composition and the composition gradient at a known depth into the object via compositional measurements of the coma.\footnote{\textit{In situ} sampling would also be an effective way to measure these values.} Assuming that the effects of cosmic ray processing are well known, this information mathematically constrains the initial composition and age of the object in most circumstances (see Sec.~\ref{sec:CRproc}). In fact, for sufficiently many observed chemical species, the age and initial composition are overconstrained and additional free parameters can be added to any model.

Sec.~\ref{sec:simpnet} then discusses a simple chemical network in which CO is processed into CO$_2$ and vice versa. While this network is overly simplified and ignores many important effects, it is sufficient to demonstrate the mathematics of this methodology and provide an example. Finally, this simple network is used to approximate the age of 3I/ATLAS (Sec.~\ref{sec:application}). This method suggests that 3I/ATLAS had an initial CO/CO$_2$ ratio of $\sim\num{1.5}$ and is older than \qty{0.6}{Gyr}. While these constraints are very loose, this application is a proof of concept for how this method can be applied and of its potential usefulness. While this CO/CO$_2$ ratio is consistent with solar system comets (e.g., \citealt{Bockelee-Morvan2017}), it also implies that 3I/ATLAS formed beyond the CO ice line, consistent with previous results \citep{Cordiner2026}. 

\subsection{Discussion}

There are three principal reasons for the imprecision of this measurement---(i) uncertainty in the effects of GCR processing, (ii) heterogeneous and distinct CO/CO$_2$ measurements, and (iii) the age and hyperactivity of 3I/ATLAS. First, the application of this method to 3I/ATLAS uses the simple CO$\leftrightarrow$CO$_2$ network. Although this process is dominant, this simplification ignores several other reactions. The behavior of these reactions can also depend on the beam used in laboratory experiments. Correcting for the beam to reproduce the impact of GCRs is non-trivial, and additional experimentation is necessary to calibrate such a model. The use of a more complex chemical and physical network---e.g., implementing a thermal model \citep{Yaginuma2026} and a more detailed chemical network such as PROCODA \citep{Pilling2022, Pilling2024}---would improve the accuracy of these models. Using more molecules would also allow this methodology to constrain additional parameters beyond the age and composition, such as the object's size and various thermal parameters. 

Second, the measurements of CO and CO$_2$ in the coma of 3I/ATLAS are highly heterogeneous, being made with different instruments, reduction methods, and times. Even when measurements are at similar epochs, there is significant scatter. With so few data points, it is difficult to know whether this scatter is due to variable emission from the coma or the different measurement techniques. The widely spaced epochs also make this method difficult to use, since the gradient has likely changed between observations. While a radius-loss model combined with the GCR processing model can approximately fit these data (see Fig.~\ref{fig:coco2ratio}), the uncertainties and scatter in these measurements make extraction difficult. For future interstellar objects, joint measurement of the CO and CO$_2$ ratio along with dust and gas mass loss rates with reasonable cadence over a relatively brief period with a single instrument would provide more precise estimates of the age and initial composition of the interstellar object.

3I/ATLAS is also not particularly well suited for this methodology. Due to its hyperactivity, the post-perihelion activity likely reflects the initial composition, making this method unnecessary. The high age of 3I/ATLAS (measured through the isotopic ratio, which is only possible for such a bright object) means that the surface is close to the steady-state composition. As a result, the composition of 3I/ATLAS' coma likely only changed during its perihelion passage (which is supported by the fit curves; see Fig.~\ref{fig:coco2ratio}). For such an object, measurement during perihelion is necessary to constrain the object's properties using this method. Since only a single data point exists in this region, it is difficult to extract this information and only a lower limit on the age can be obtained. However, these constraints on the applicability of this method are not a major concern. If a future interstellar object is similarly hyperactive, it will likely be sufficiently bright for isotopic age measurements to be successful. For such an object, even low-quality measurements of the CO/CO$_2$ ratio, when combined with the age, will constrain the initial composition. 

There are some caveats that must be noted about this method of measuring interstellar object properties. For one, this method only works on comet-like objects with significant comae. An asteroidal or point-source interstellar object, like 1I/`Oumuamua, is not a suitable target. This method also is highly sensitive to the GCR processing model. A strong understanding of the effects of GCRs is therefore critical for accurate and precise constraints, which requires additional laboratory experiments. This model also relies on knowing the depth into the interstellar object where the composition is being measured. Early detection and measurement of dust and gas mass loss are critical to successfully approximating the radius loss and thus constraining the object's properties.

More critically, this method does not measure the time since the interstellar object was formed, but the time since it was ejected from its host system. Recent work used the late onset of methane emission from 3I/ATLAS, with the assumption that this onset was the result of GCR processing, to find a ``processing age'' of \qtyrange{0.9}{3.1}{Gyr} \citep{Rawat2026a}. This result is in tension with the high-precision isotopic measurement, which implies an age of \qtyrange{10}{12}{Gyr}. Although this discrepancy may be resolved by suggesting that 3I/ATLAS was ejected from its original system \qtyrange{7}{11}{Gyr} after formation, detecting such an object is unlikely. The vast majority of interstellar objects are ejected by instability and migration in young planetary systems ($\leq\qty{1}{Gyr}$), with only a small fraction remaining in the planetary system (see, e.g., \citealt{Tremaine1993, Wyatt2017, Levine2023} and references therein). Although these remaining objects are constantly ejected, the early-ejected population dominates. It is therefore highly unlikely for the ``processing age'' of an interstellar object to be significantly different from its formation age. 

The resolution of this discrepancy likely lies in the GCR processing model used by those authors, which does not include the CO$_2$$\rightarrow$CO reaction. This exclusion means that GCR processing eventually destroys all surface CO and CO$_2$ in their model, forcing the ``processing age'' to be smaller than if the back-reaction were included. Regardless of these comments, it remains true that the age inferred from the GCR processing as discussed here may be different from the true age of the interstellar object, either due to late ejection or variable GCR fluxes during the interstellar object's time in the interstellar medium.

There are also other potential causes for the changing CO/CO$_2$ ratio of 3I/ATLAS beyond GCR processing. Since CO and CO$_2$ sublimate at different temperatures, they are expected to activate at different times and have different outgassing rates during perihelion passage. While taking the ratio of CO and CO$_2$ partially corrects for these effects, part of the changing CO/CO$_2$ ratio can be ascribed to the different sublimation temperatures. Although erosion from the sublimation of a single dominant species drives most of the mass loss (cf. the $<\qty{1}{m}$ sublimation depths of \citealt{Yaginuma2026} and the tens of meters of erosion found here), this effect can be fully accounted for by a thermal model. 

Another possible confounder is nuclear heterogeneity, as patches of CO and CO$_2$ can be exposed as the nucleus erodes. If the sampled composition changes are primordial in this way, then the composition changes are independent of GCR processing and are independent of the age. Fortunately, the time-series coma monitoring that is ideal for this technique would also detect the rapid changes in coma abundance that would indicate such heterogeneous structure. 

Future work is needed to improve the accuracy of these models, develop useful observational strategies, and constrain the ages and compositions of interstellar objects. As the Vera C. Rubin Observatory's Legacy Survey of Space and Time finds more of these objects, this method will be a powerful tool to investigate the population of interstellar objects as a whole and individually. The age and composition distributions of this population will provide a unique insight into planet formation in extrasolar systems. 

\section*{Acknowledgments}

I thank Fred Adams, Adam Distler, Tessa Frincke, Preston Kemeny, Luis Salazar Manzano, Darryl Seligman, and James Sunseri for insightful conversations. A.G.B.T. acknowledges support from the Fannie and John Hertz Foundation and the University of Michigan's Rackham Merit Fellowship Program. 

Claude Code (Sonnet/Opus 5; Anthropic) was used to help write code, create figures, and edit manuscript prose. The author takes full responsibility for its contents. 

\appendix
\vspace{-20pt}
\section{Negativity of Network Eigenvalues}\label{app:negeigproof}

In this Section, I prove that the rate matrix $\mathbf{R}$'s eigenvalues all have a real part less than or equal to zero, so that the system can only decay or be in a steady state. 

Consider the matrix $\mathbf{A}$, which is irreducible, has the property that $A_{ii}=-\sum_{j\neq i}A_{ji}$, and is a Metzler matrix (so that $A_{ji}\geq0$ for $i\neq j$). The vector $\vec{1}^\top$ is therefore a left-eigenvector of $\mathbf{A}$ with eigenvalue zero. All other eigenvalues of this matrix are guaranteed to have a negative real part. 

This can be shown by taking a new matrix $\mathbf{M}=\mathbf{A}+c\mathbf{I}$, where $\mathbf{I}$ is the identity matrix and $c$ is the negative of the largest-magnitude number on the diagonal of $\mathbf{A}$. By construction, $\mathbf{M}$ is nonnegative, and the eigenvalues $\mu$ of $\mathbf{M}$ are related to the eigenvalues $\mathbf{\lambda}$ of $\mathbf{A}$ by $\lambda=\mu-c$.

By the Perron-Frobenius theorem \citep{Seneta1981}, the (irreducible) matrix $\mathbf{M}$ has exactly one eigenvector with all positive components, and this eigenvector corresponds to a maximal eigenvalue. This eigenvector is $\vec{1}$, which is a left-eigenvector with an eigenvalue of $c$. By Perron-Frobenius, $|\mu|<c$ for all other $\mu$. 

Substituting $\mu=\lambda+c$ and $\lambda=a+bi$, this relation becomes 
\begin{equation}
    a<-\frac{a^2+b^2}{2c}\,.
\end{equation}
Since $a^2$, $b^2$, and $c$ are non-negative, $a=\Re[\lambda]\leq0$ for all $\lambda$. Therefore, the network described by $\mathbf{A}$ is composed of decaying components (which may rotate) and exactly one steady-state component. 

The inclusion of the $\vec{B}$ term complicates matters slightly, but can be handled. By Perron-Frobenius, since $\mathbf{A}-\text{diag}(\vec{B})\leq\mathbf{A}$ for all elements, the largest real part of any eigenvalue of $\mathbf{A}-\text{diag}(\vec{B})$ is less than or equal to that of $\mathbf{A}$. Since I have already shown that the largest eigenvalue of $\mathbf{A}$ is zero, the inclusion of the $\vec{B}$ term can only make the system decay more rapidly. 
    
\bibliography{main}{}
\bibliographystyle{aasjournal}

\end{document}